\documentclass[12 pt,twoside,prd,amsmath,amssymb,tightenlines,showpacs,
showkeys,eqsecnum]{revtex4}
\renewcommand{\cal}{\mathcal}

\renewcommand{\hat}{\widehat}

\newcommand {\ve}{\varepsilon}
\newcommand {\pr}{\partial}
\newcommand {\cL}{\cal L}
\newcommand {\vf}{\varphi}

\begin{document}
\draft
\title{Causal structures in the four dimensional Euclidean space}
\author{Ivanhoe B. Pestov}
\affiliation{Bogoliubov Laboratory of Theoretical Physics\\ Joint
Institute for Nuclear Research\\ 141980 Dubna, Moscow region,
Russia} \email{pestov@thsun1.jinr.ru}
\author{Bijan Saha}
\affiliation{Laboratory of Information Technologies\\ Joint
Institute for Nuclear Research\\ 141980 Dubna, Moscow region,
Russia} \email{saha@thsun1.jinr.ru}
\homepage{http://thsun1.jinr.ru/~saha}
\date{\today }

\begin{abstract}

It is  shown that in the 4d Euclidean space there are two causal
structures defined by the temporal field. One of them is
well-known Minkowski spacetime. In this case the gravitational
potential (the positive definite Riemann metric) and temporal
field satisfy the Einstein equations with trivial energy-momentum
tensor. However, in the case of the second causal structure the
gravitational potential and temporal field should be connected
with some nontrivial energy-momentum tensor. We consider the
simplest case with energy-momentum tensor of the real scalar field
and derive exact solution of the field equations. It can be viewed
as the ground to consider the second causal structure on the equal
footing with the Minkowski spacetime, i.e., as an object
interesting from the physical point of view, especially in the
framework of the field theory.

\end{abstract}


\maketitle

\section{Introduction}

Einstein \cite{1} set up the problem of including gravity into the
framework of the
Faraday concept of field and formulated the principles of gravitational
physics (General Theory of Relativity).
However, it turned out that all important consequences of the theory  were
not generally covariant. The reason for the appearance of
noncovariant results in the
theory with the principle of general covariance as a cornerstone  of all
building is still unclear. The problems turned out to be so
serious  that one begins to gradually consider
them as  verifications of the exceptional properties of the gravitational
field such as universality, nonlocality and nonscreening.
A variety of  examples of the discrepancy between the first principles of
General Relativity  and the results that follow from it
can be found, for example, in
~\cite{2,3,4,5,6}. This includes  all kinds of the so-called pseudo-energy
tensor of the gravitational field, the background metrics introduced in one
form or another and alike.

In ~\cite{7} one of the authors (I.B.P.) has given a simple solution of
this actual and paradoxical problem in question. The essence of
new approach is as follows. Einstein placed in the foundation of
General Relativity
the principle of equivalence and tightly connected with it the principle
of general covariance [1],[3].  The principle of general
covariance has an intimate physical meaning,
which in particular implies that coordinates (that one uses for description
of physical phenomena) have neither physical nor geometric significance. They
are simply parameters. From this follows an evident and exact
requirement
to formulate general covariant theory of the gravitational field without
using the well-known  representations about space
and time. Only in the case when one
is able to analyze how time and space emerge in the framework of the
general covariance principles, it is possible to provide answers to such a
key problem as essence of time and space.
A principal result of the analysis underlies in the fact that
time is not an additional dimension but a field ~\cite{7}. The field concept
of time
discloses the reason behind those difficulties that exist in  General
Relativity  and removing them it uncovers General Relativity as the
self-consistent physical theory in which energy is as fundamental
as in  quantum mechanics.

The notion of smooth manifold is a geometric basis of General
Relativity because this structure does not distinguish
intrinsically between different coordinate systems (the principle
of general covariance is naturally included into this notion). A
manifold is not given apriori (as it is usually presupposed) but
is defined by the physical system.  It was shown ~\cite{7}, that
to this end one needs to put in correspondence to the
gravitational field a positive definite Riemann metrics, i.e., the
field that was first introduced   by Riemann in the framework of
geometrical concepts ~\cite{3,6}. A smooth four dimensional
manifold that corresponds to a physical system is called a
physical manifold.

In this paper we consider a four-dimensional Euclidean space as
physical manifold and hence gravitational potential is known. By
virtue of it the main equation for the temporal field can be
exactly solved.  We consider two exact solutions to this equation.
One of them is invariant with respect to the translation along the
stream of time and other one is invariant with respect to the
rotations. In the Sect. II we demonstrate that the first solution
defines the causal structure, which is known as the Minkowski
spacetime. In Sect. III the second causal structure, defined by
the second solution, is considered. In Sect. IV we consider a
system of gravitational, temporal and real scalar fields and give
exact solutions to the corresponding equations. This enables us to
find important physical quantities such as energy density, energy
flow vector of the gravitational and scalar fields, respectively,
and total energy of the system in question. On this ground one can
consider the second causal structure (like the Minkowski
spacetime) as an object interesting from the physical point of
view.

\section{Usual causal structure}
  The vector space $R^4$ is the set of 4-tuples of real numbers. The symbol
  $\stackrel{\rightarrow}{u} $ denotes the vector in $R^4$  with components
  $(u^1, \, u^2, \, u^3, \, u^4)$ ; $\stackrel{\rightarrow}{u}  \cdot
  \stackrel{\rightarrow}{v} $  denotes the usual scalar product
  $\stackrel{\rightarrow}{u}  \cdot    \stackrel{\rightarrow}{v}: =
  u^1 v^1 + u^2 v^2 + u^3 v^3 + u^4 v^4 .$  The distance function be
  $$d(\stackrel{\rightarrow}{u}, \stackrel{\rightarrow}{v}  ) =
  \sqrt {(\stackrel{\rightarrow}{u} - \stackrel{\rightarrow}{v}) \cdot
  (\stackrel{\rightarrow}{u} - \stackrel{\rightarrow}{v})    }, \quad
  d(\stackrel{\rightarrow}{u}, 0)  =
  |\stackrel{\rightarrow}{u}|.$$
  In accordance with [7], the causal structure on $R^4$  is defined
  by the metric $g_{ij} = \delta_{ij}$ and the scalar field
  $f(u^1, \, u^2, \, u^3, \, u^4 )$ (field of time or temporal field), which
  is  the solution to the equation
  \begin{equation}
  \biggl(\frac{\partial f}{\partial u^1} \biggr)^2 +
  \biggl(\frac{\partial f}{\partial u^2} \biggr)^2 +
  \biggl(\frac{\partial f}{\partial u^3} \biggr)^2 +
  \biggl(\frac{\partial f}{\partial u^4} \biggr)^2 = 1.
  \label{1}
  \end{equation}
  Space cross-section of the $R^4$ is defined by the field of time. For
  a given real number $t$, it is given by the equation
  $ f(u^1, \, u^2, \, u^3, \, u^4 )=t. $
  It is evident that the equation (\ref{1}) has the solution
  $ f(u^1, \, u^2, \, u^3, \, u^4 )
  =  \stackrel{\rightarrow}{a} \cdot \stackrel{\rightarrow}{u},$
  where $\stackrel{\rightarrow}{a}= (a^1, \, a^2, \, a^3, \, a^4) $  is a
  unit constant vector
  $\stackrel{\rightarrow}{a} \cdot \stackrel{\rightarrow}{a} =1.$
 The stream of time is defined in the case in question as a congruence
 of lines of time on
 the physical manifold $R^4.$ Analytically,
 the lines of time are defined as a solution of the autonomous
 system of differential equations
 $$  \frac{d u^i}{dt} = g^{ij} \frac{\partial f}{\partial u^j} = t^i =a^i.$$
The general solution is a straight line that goes through the fixed point
$ \stackrel{\rightarrow}{u_0}:  $
\begin{equation}
\stackrel{\rightarrow}{u}(t) = \stackrel{\rightarrow}{a}(t-t_0) +
 \stackrel{\rightarrow}{u_0}. \label{2}
\end{equation}
 The  causal structure in question defines the interval as follows. Let
 $$\stackrel{\rightarrow}{u}_{s} = \stackrel{\rightarrow}{u} -2
 \stackrel{\rightarrow}{a} (\stackrel{\rightarrow}{a} \cdot
 \stackrel{\rightarrow}{u}) $$ be the vector symmetrical
  $\stackrel{\rightarrow}{u}   $  with respect to the vector
   $ \stackrel{\rightarrow}{a} . $  Then in the coordinates
   $u^1, \, u^2, \, u^3, \, u^4 $ the interval can be presented as follows:
   $$ s^2 =\stackrel{\rightarrow}{u}\cdot \stackrel{\rightarrow}{u}_{s}
   =\stackrel{\rightarrow}{u}\cdot \stackrel{\rightarrow}{u} -
   2 (\stackrel{\rightarrow}{a}\cdot \stackrel{\rightarrow}{u}  )^2
   = -|\stackrel{\rightarrow}{u}|
   \cos2{\theta} , $$
where $\theta $  is an angle between $\stackrel{\rightarrow}{a} $
and $\stackrel{\rightarrow}{u}.$ To be sure that $s$ is really well-known
interval,
let us introduce the system of coordinates compatible with
causal structure [7]. To this end suppose that all initial data in (\ref{2})
belong to the space-section
\begin{equation}
\stackrel{\rightarrow}{a} \cdot \stackrel{\rightarrow}{u}_{0}= t_0,
\label{3}
\end{equation}
(the space $R^3$  orthogonal to the vector $\stackrel{\rightarrow}{a}$.)

Let us consider the natural system  of four orthogonal unit vectors
$$\stackrel{\rightarrow}{E}_{0} =(a^1,\, a^2,\, a^3,\, a^4 ), \quad
\stackrel{\rightarrow}{E}_{1} =(-a^4,\, -a^3,\, a^2,\, a^1 ) ,  $$
$$\stackrel{\rightarrow}{E}_{2} =(a^3,\, -a^4,\, -a^1,\, a^2 ), \quad
 \stackrel{\rightarrow}{E}_{3} =(-a^2,\, a^1,\, -a^4,\, a^3 ), $$
$\stackrel{\rightarrow}{E}_{0} = \stackrel{\rightarrow}{a}.  $
Now the general solution to equation (\ref{3})  has the form
$$ \stackrel{\rightarrow}{u}_{0} = t_0 \stackrel{\rightarrow}{E}_{0} +
 x  \stackrel{\rightarrow}{E}_{1} + y \stackrel{\rightarrow}{E}_{2} +
z \stackrel{\rightarrow}{E}_{3} .$$
Substituting this representation into
 formula (\ref{3}) we obtain that the congruence of the lines of time
 can be written in the following form:
 \begin{equation}
\stackrel{\rightarrow}{u} = t \stackrel{\rightarrow}{E}_{0} +
x  \stackrel{\rightarrow}{E}_{1} + y \stackrel{\rightarrow}{E}_{2} +
z \stackrel{\rightarrow}{E}_{3}.\label{4}
\end{equation}
We can consider (\ref{4}) as the coordinate transformation (with unit
Jacobean). It is easy to see that in the coordinates $t, \, x,\, y,\,z,$
the interval takes the form:
$$ s^2 = x^2 + y^2 + z^2 -t^2 . $$
Let us consider one more example. If we put
  $${\hat P}_0 = \stackrel{\rightarrow}{E}_{0} \cdot
  \stackrel{\rightarrow}{\nabla}, \quad  {\hat P}_1 =
  \stackrel{\rightarrow}{E}_{1} \cdot \stackrel{\rightarrow}{\nabla}, \quad
   {\hat P}_2  = \stackrel{\rightarrow}{E}_{2}
  \cdot \stackrel{\rightarrow}{\nabla}, \quad   {\hat P}_3  =
   \stackrel{\rightarrow}{E}_{3} \cdot \stackrel{\rightarrow}{\nabla}, $$
   where $$\stackrel{\rightarrow}{\nabla} =
   (\frac{\partial}{\partial u^1}, \, \frac{\partial}{\partial u^2}, \,
   \frac{\partial}{\partial u^3}, \, \frac{\partial}{\partial u^4} ),$$
   then the Dirac equation in the coordinates $u^1, \, u^2, \, u^3, \, u^4 $
   reads
   $$ i \gamma^{\mu} {\hat P}_{\mu}\psi  = \frac{mc}{\hbar} \psi.$$
The Dirac equation in the coordinates $t, \, x,\, y,\,z$
has a ordinary form

$$ i( {\gamma}^{0} \frac{\partial}{\partial t} +
{\gamma}^{1} \frac{\partial}{\partial x} +
{\gamma}^{2} \frac{\partial}{\partial y} +
{\gamma}^{3} \frac{\partial}{\partial z}) \psi  =
\frac{mc}{\hbar} \psi.     $$
One can work in either the coordinates $u^1, \, u^2, \, u^3, \, u^4 $  or
in the coordinates $t, \, x,\, y,\,z,$  but in
the first case the physical
results should not depend on the choice of the constant vector
$\stackrel{\rightarrow}{a} $ (a direction of time flux).

\section{ The new causal structure}

As it is shown, the considered causal structure  on $R^4$
 is the Minkowski space-time. Now we shall show that there is a new
 causal structure on $R^4$ that is invariant with respect to the
 rotations. Indeed, equation (\ref{1}) has a remarkable solution
 \begin{equation}
 f(u^1, \, u^2, \, u^3, \, u^4 )= \sqrt{(\stackrel{\rightarrow}{u}  \cdot
 \stackrel{\rightarrow}{u} )} =
         \sqrt{ (u^1)^2 + (u^2)^2 + (u^3)^2 + (u^4)^2}.
         \label{5}
\end{equation}
 Thus, in the first case the space sections of the
 $R^4$ are $R^3$
 ( $ \stackrel{\rightarrow}{a} \cdot \stackrel{\rightarrow}{u}=t $ )
 and in the second one the space sections are $3D$  spheres   $S^3$
 $ |\stackrel{\rightarrow}{u}|
  =\sqrt{(\stackrel{\rightarrow}{u} \cdot \stackrel{\rightarrow}{u} )} =
 \tau. $ In this case, the lines of time  are the solutions of the system of
 equations
 $$  \frac{du^i}{d\tau} =  \frac{\partial f}{\partial u^i} =
 \frac{u^i}{f} = \frac{u^i}{\sqrt{ (u^1)^2 + (u^2)^2 + (u^3)^2 + (u^4)^2}}. $$
 It can be shown that the general solution of this system is
 \begin{equation}
 u^i(\tau) = {u^i_0}\frac{\tau}{\tau_0},    \label{3.2}
 \end{equation}
 during which the initial data belong to the space-section
 \begin{equation}
 \stackrel{\rightarrow}{u}_{0} \cdot \stackrel{\rightarrow}{u}_{0} =
 {\tau_0}^2.   \label{3.3}
  \end{equation}
   We see that the second causal structure is the  congruence
 of rays coming from initial point $(0,\,0,\,0,\,0)$ of $R^4 $ (origin).
 Thus, the coordinate space $R^4$ can be fibered on  $R^3$ (the stream of
 time is the bundle of parallel straight lines) or $S^3$ (the stream of
 time is the bundle of the rays coming from one point).

 The action for the point particle defined by the second causal structure
 has the form
 $$ S = -mc \int \limits_{p}^{q} \sqrt{1- {\tau}^2 {\omega}^2} d\tau , $$
 where $\omega = dl / d\tau$ and $dl$   is the element of arc on the unit
 $3D$ sphere.  Really, $ \stackrel{\rightarrow}{du}  \cdot
\stackrel{\rightarrow}{du} = d\tau^2 + \tau^2 dl^2,$    and
$\stackrel{\rightarrow}{u} \cdot  \stackrel{\rightarrow}{du} =
\tau d\tau.$ Let us introduce the system of coordinates compatible
with causal structure. Rewrite the relation (\ref{3.3}) in the
parametric form
\begin{eqnarray}
u_0^1 &=& \tau_0 \sin{\alpha}\sin{\vartheta}\cos{\phi}, \nonumber\\
u_0^2 &=& \tau_0 \sin{\alpha}\sin{\vartheta}\sin{\phi}, \nonumber\\
u_0^3 &=& \tau_0 \sin{\alpha}\cos{\vartheta},\nonumber\\
u_0^4 &=& \tau_0 \cos{\alpha}, \nonumber
\end{eqnarray}
where $(0 \le \alpha, \vartheta \le \pi,\, 0 \le \phi \le 2 \pi)$.
Inserting this representation into (\ref{3.2}) we come to the
conclusion that the congruence of the lines of time can be written
as follows:
\begin{subequations}
\label{01}
\begin{eqnarray}
u^1 &=& \tau \sin{\alpha}\sin{\vartheta}\cos{\phi}, \label{01a}\\
u^2 &=& \tau \sin{\alpha}\sin{\vartheta}\sin{\phi},\label{01b}\\
u^3 &=& \tau \sin{\alpha}\cos{\vartheta},\label{01c}\\ u^4 &=&
\tau \cos{\alpha}. \label{01d}
\end{eqnarray}
\end{subequations}
We can consider the transformation (\ref{01}) as the coordinate
transformation. In the coordinate $(\tau,\, \alpha,\,\vartheta,\,\phi)$
we have the metric
\begin{equation}
ds^2 = d \tau^2 + \tau^2 \bigl[d\alpha^2 + \sin^2 \alpha \bigl( d
\vartheta^2 + \sin^2 \vartheta d \phi^2\bigr)\bigr],
\label{metric}
\end{equation}
with
\begin{equation}
\sqrt{g} = \tau^3 \sin^2 \alpha \sin \vartheta
\label{sqg}
\end{equation}
and gradient of temporal field has the form $t^i =(0,\,0,\,0,\,1).$

\section{Exact solution of the field equations}

In this section we study the self-consistent system of the gravitational,
temporal and real scalar field equations for the reason to follow. The
geometrical equations of the gravitational field read ~\cite{7}
\begin{equation}
G_{ij} + T_{ij} = \ve t_i t_j,  \quad \ve = \ve_m + \ve_g,
\label{einmod}
\end{equation}
where $\ve_m$ and $\ve_g$ are the energy densities of the matter
and gravitational field, respectively and given by
\begin{subequations}
\begin{eqnarray}
\ve_m &=& T_{ij}t^it^j, \label{edmat}\\
\ve_g &=& G_{ij} t^it^j.\label{edgr}
\end{eqnarray}
\end{subequations}
Here
$${G}_{ij} = \tilde{R}_{ij} - \frac{1}{2} \tilde{g}_{ij} \tilde{R}                $$
is the Einstein tensor defined by the subsidiary metric
$$ \tilde{g}_{ij} = g_{ij} - 2t_i t_j, \quad
\tilde{g}^{ij} = g^{ij} - 2t^i t^j.$$
Note that quantities like $R$ and $R_{ij}$ are related to the metric
$g_{ij}$ while those with tilde, i.e., $\tilde{R}$ and
$\tilde{R}_{ij}$ are generated from the subsidiary metric $\tilde{g}_{ij}$;
$t^i$ is the gradient of temporal field defined as
$$t^i = g^{ij}\frac{\partial f}{\partial u^i}.$$
It is easy to see that the first causal structure defined by
the $g_{ij} = \delta_{ij}$ and $f = a_i u^i$ is the solution to
the Eq. (\ref{einmod}) with $T_{ij} = 0$ and in this case $\ve_g = 0.$
But for the second causal structure defined by
the $g_{ij} = \delta_{ij}$ and $f = \sqrt{u_i u^i}$ is not the solution to
the Eq. (\ref{einmod}) with $T_{ij} = 0$. We see that second causal
structure will be physical field under the condition that the
energy-momentum tensor is not trivial. We shall give exact solution
of this problem for the case with energy-momentum tensor of real
scalar field.

The scalar field Lagrangian we choose in the form:
\begin{equation}
{\cL}_{\vf} = \frac{1}{2} \tilde{g}^{ij} \pr_i \vf \pr_j \vf + V(\vf). \label{sclag}
\end{equation}
Here $V(\vf)$ is some unknown potential energy that should be fixed as
a result of the solution of Eq. (\ref{einmod}).
Taking into account that
\begin{equation}
\tilde{R} = \frac{12}{f^2}, \qquad \tilde{R}_{ij} =
\frac{4}{f^2} (g_{ij} - t_i t_j),
\label{Riccidef}
\end{equation}
for the Einstein tensor one finds
\begin{equation}
{G}_{ij} =
- \frac{2}{f^2} g_{ij} + \frac{8}{f^2} t_i t_j.
\label{eint}
\end{equation}
Now in account of the fact that in our case both
$R$ and $R_{ij}$ are trivial, from (\ref{edgr}) we find
\begin{equation}
\ve_g = 6/f^2.  \label{veg}
\end{equation}
In accordance with (\ref{eint}) and (\ref{veg}) we consider the case
when $\vf = \vf (f)$ with $f = f(u) = \sqrt{u_iu^i} =
\sqrt{u_1u^1 + u_2u^2 +u_3u^3 +u_4u^4}.$
On account of the energy-momentum tensor $T_{ij}$
\begin{equation}
T_{ij} = \pr_i \vf \pr_j \vf - \tilde{g}^{ij} {\cL}_{\vf}, \label{emtsc}
\end{equation}
we find the energy density of the scalar field as
\begin{equation}
\ve_\vf = T_{ij}t^it^j = \frac{1}{2} {\vf^\prime}^2 + V(\vf), \label{edsc1}
\end{equation}
whereas, for the energy flow vector of the scalar field we find
\begin{equation}
\Pi_i = \ve_\vf t_i - T_{ij} t^j = 0. \label{moment}
\end{equation}
In view of what has been told, from (\ref{einmod})
we find
\begin{equation}
\frac{1}{2} {\vf^\prime}^2 - V(\vf) - \frac{2}{f^2} = 0.
\label{einmodvf}
\end{equation}
On the other hand, the equation of motion
\begin{equation}
\frac{1}{\sqrt{g}} \pr_i (\sqrt{g} g^{ij} \pr_j \vf) -
\frac{d V(\vf)}{d \vf} = 0,
\label{eom}
\end{equation}
gives
\begin{equation}
\vf^{\prime \prime} + \frac{3}{f} \vf^\prime + \frac{d V(\vf)}{d \vf} = 0.
\label{eom1}
\end{equation}
Excluding $\vf$ from (\ref{einmodvf}) and (\ref{eom1}) we find
the second order differential equation for $\vf$:
\begin{equation}
2\vf^\prime \vf^{\prime \prime} + \frac{3 \vf^{\prime 2}}{f}
+ \frac{4}{f^3} = 0.
\label{em1}
\end{equation}
The Eq. (\ref{em1}) allows the following first integral:
\begin{equation}
{\vf^\prime}^2 = \frac{4L}{f^3} - \frac{4}{f^2}.
\label{vfpr}
\end{equation}
Here $L$ is the constant of integration. From Eq. (\ref{vfpr})
it follows that the physical manifold is not $R^4$ but a ball
$f^2 \le L^2$. For $\vf$ finally we find
\begin{equation}
\vf = \mp 4 \Biggl[\sqrt{\frac{L}{f} - 1}\,\, +\,\,
\arcsin{\sqrt{\frac{f}{L}}}\Biggr] \pm 2\pi,
\label{vfsol}
\end{equation}
where the integration constant $C = 2 \pi$ is defined from the
condition $\vf(L) = 0$.
The corresponding potential now can be found from (\ref{einmodvf}):
\begin{equation}
V(\vf) = \frac{2L}{f^3} - \frac{4}{f^2}. \label{potexpl}
\end{equation}
Using the results obtained we find the energy density of the scalar,
temporal and gravitational field system.
It gives
\begin{eqnarray}
\ve = \ve_\vf + \ve_g = \frac{\vf^{\prime 2}}{2} +V(\vf) + \frac{6}{f^2}
= \frac{2L}{f^3}-\frac{2}{f^2}+\frac{2L}{f^3}-\frac{4}{f^2}+
\frac{6}{f^2} = \frac{4L}{f^3}.
\label{totend}
\end{eqnarray}
 The law of energy conservation reads ~\cite{7}:
 $$   \nabla_i (\ve  t^i) =0. $$
 In our case
 $$ \nabla_i = \frac{\partial}{\partial u^i}, \quad \ve = \frac{4L}{f^3},
 \quad t^i = \frac{u^i}{f}$$ and hence
 $$\nabla_i (\ve  t^i) = 4L \frac{\partial}{\partial u^i} (\frac{u^i}{f^4})=
 4L(\frac{4}{f^4} - \frac{4}{f^5}u^i \frac{\partial}{\partial u^i} f ) =0. $$
 In the system of coordinates compatible with causal structure we should have
 $$ \frac{\partial}{\partial \tau}(\sqrt{g} \ve) =0       $$
 and this is the case because from (\ref{sqg}) and (\ref{totend}) it follows that
 $$ \sqrt{g} \ve =  \tau^3 \sin^2 \alpha \sin \vartheta \cdot
 \frac{4L}{\tau^3} =4L \sin^2 \alpha \sin \vartheta .    $$
 If we integrate with respect to the variable
 $ \alpha,\,  \vartheta,\,\phi   $   we find
 for the total energy of the system
\begin{equation}
E = 8 \pi^2 L.
\label{toen}
\end{equation}
We see that the solution obtained leads to the reasonable result.
It can be shown also that the energy flow vector of the gravitational field is
trivial.

Thus, we found exact solution of the corresponding equations. With this
we found all important physical quantities: the energy density,
an energy flow vector of the gravitational and scalar field and
total energy of the system in question.

\end{document}